\documentstyle[epsfig,12pt]{article}  

\newcommand{\beq}{\begin{equation}}
\newcommand{\eeq}{\end{equation}}
\newcommand{\bea}{\vspace{0.25cm}\begin{eqnarray}}
\newcommand{\eea}{\end{eqnarray}}
\newcommand{\r}{\mbox{{\boldmath
$\rho$}}}

\newcommand{\pb}{\mbox{{\bf
p}}}

\newcommand{\fb}{\mbox{{\bf
f}}}

\def\lsim{\mathrel{\rlap{\lower4pt\hbox{\hskip1pt$\sim$}}
    \raise1pt\hbox{$<$}}}         %less than or approx. symbol
\def\gsim{\mathrel{\rlap{\lower4pt\hbox{\hskip1pt$\sim$}}
    \raise1pt\hbox{$>$}}}         %greater than or approx. symbol
\begin{document}
\thispagestyle{empty}
\vspace*{-2cm}
 
\bigskip
 
\begin{center}

  {\large\bf
EFFECT OF ELECTRIC FIELD OF THE ELECTROSPHERE ON PHOTON
EMISSION FROM QUARK STARS
\\
\vspace{1.5cm}
  }
\medskip
  {\large
B.G. Zakharov
}
  \bigskip

{\it
L.D. Landau Institute for Theoretical Physics,
        GSP-1, 117940,\\ Kosygina Str. 2, 117334 Moscow, Russia
\vspace{2.7cm}\\}
  {\bf
  Abstract}
\end{center}
{
\baselineskip=9pt
We investigate the photon emission from the electrosphere
of a quark star. It is shown that at temperatures $T\sim 0.1\div 1$ MeV
the dominating mechanism is the bremsstrahlung due 
to bending of electron trajectories in the mean Coulomb field of the
electrosphere. The  radiated energy flux from this mechanism
exceeds considerably both the contribution from the bremsstrahlung 
due to electron-electron
interaction and the tunnel $e^{+}e^{-}$ pair creation.\\
}
\vspace{1cm}

%\noindent{\sl Keywords: photon emission, relativistic 
%plasma, quark stars}
%-------------------------------------------------------------
\newpage
%-------------------------------------------------------------

\noindent {\bf 1}.
It is possible that quark stars 
made of a stable strange quark matter (SQM) \cite{Witten,Alcock1,Haensel}
(if it exists) may exist without a crust of normal matter \cite{Usov1}.
The quark density for bare quark stars should drop abruptly at 
the scale $\sim 1$ fm.
The SQM in normal phase and in the two-flavor superconducting  (2SC) phase 
should also contain electrons 
(for normal phase the electron chemical potential, $\mu$, 
is about 
20 MeV \cite{Alcock1,Kettner}). 
Contrary to the quark density the 
electron density  drops smoothly above the star surface at the 
scale $\sim 10^{3}$ fm \cite{Alcock1,Kettner}. 
For the star surface temperature 
$T\ll \mu$, say $T\lsim 10^{10} \mbox{K}\sim 1$ MeV, 
this ``electron atmosphere'' (usually called the electrosphere) 
may be viewed as 
a strongly degenerate relativistic
electron gas \cite{Alcock1,Kettner}. 
The photon emission from the normal SQM is negligibly small as compared
to the black body one at $T\ll \omega_{p}$ \cite{Chmaj,Harko1} (here 
$\omega_{p}\sim 20$ MeV  is the plasma frequency of the SQM
\cite{Chmaj}).
However, for the electrosphere the plasma frequency is much smaller
than that for the SQM. For this reason the photon emission from 
the electrosphere may potentially dominate the luminosity of a quark 
star. 
Contrary to neutron stars 
(or quark stars with a crust of normal matter)
the photon emission from the electrosphere of bare quark stars
may exceeds the Eddington limit, and may be used for 
distinguishing a bare quark star from a neutron star (or a quark 
star with a crust of normal matter). For this reason it is of great 
importance to have 
quantitative predictions for the photon emission from the electrosphere.
This is also of interest in the context of the scenario of the gamma-ray 
repeaters due to reheating of a quark star by impact of a massive comet-like
object \cite{Usov-GRB}.

The bremsstrahlung from the electrosphere due to the 
electron-electron interaction has been addressed
in \cite{Gale,Zh1}. 
The authors of \cite{Gale} used the soft photon approximation
and factorized the $e+e\to e+e$ cross section
in the spirit of Low's theorem.
In \cite{Zh1} it was pointed out that this approximation is inadequate
since it neglects the effect of the photon energy on the electron 
Pauli-blocking which should lead to a strong overestimate of the 
radiation rate. The authors of \cite{Zh1} have not given a consistent
treatment of this problem either. To take into account the effect of the
minimal photon energy they suggested some restrictions on the initial electron
momenta introduced by hand.
In this way they obtained the 
radiated energy flux 
from the
$e^{-}e^{-}\to e^{-}e^{-}\gamma$ process
which is much smaller than 
that in \cite{Gale}, and than the energy flux
from annihilation of positrons produced in
the tunnel $e^{+}e^{-}$ creation in the electric field of the electrosphere 
\cite{Usov1,Usov2}. 
In \cite{Harko2} there was an attempt to include the effect of the mean Coulomb
field of the electrosphere on the photon emission. The authors obtained
a considerable enhancement of the radiation rate.
However, similarly to \cite{Gale} the analysis
\cite{Harko2} treats incorrectly the Pauli-blocking effect.

Thus the theoretical situation with 
the photon bremsstrahlung from the electrosphere is still controversial 
and uncertain. The main problem here is an accurate accounting for 
the photon energy in the Pauli-blocking.
In the present paper we address the bremsstrahlung
from the electrosphere in a way similar to the Arnold-Moore-Yaffe
(AMY) \cite{AMY1}  approach to the collinear photon emission from 
a hot quark-gluon plasma  within the thermal field theory.
We use a reformulation of the AMY formalism given in \cite{AZ}
which is based on the light-cone path integral (LCPI) 
approach \cite{Z1,Z_SLAC1,Z_SLAC2} (for reviews, see 
\cite{Z_YAF98,Z_NP2005}) to the radiation processes. 
For an infinite homogeneous plasma 
(with zero mean field) 
the formalism  \cite{AZ} reproduces the 
AMY results \cite{AMY1}.
The LCPI formulation \cite{AZ} has the advantage that it also works for 
plasmas with nonzero mean field. 
It allows to evaluate the photon emission accounting for 
bending of the electron trajectories in the mean Coulomb potential 
of the electrosphere.
Contrary to very crude and qualitative methods of
\cite{Gale,Zh1,Harko2} the treatment of the Pauli-blocking
effects in \cite{AMY1,AZ} has robust quantum field theoretical
grounds. 
Of course, our approach is only valid in the regime of collinear 
photon emission when the dominating photon energies exceed several 
units of the photon quasiparticle mass. Numerical calculations show 
that even at $T\sim 0.1$ MeV the effect of the noncollinear 
configurations is relatively small.  

We demonstrate that for the temperatures $T\sim 0.1\div 1$ MeV 
the radiated energy flux from the  $e^{-}\to e^{-}\gamma$ transition in 
the mean electric field  
turns out to be much bigger than contributions from the
$e^{-}e^{-}\to e^{-}e^{-}\gamma$ process
and the tunnel $e^{+}e^{-}$ creation. Our results show that the photon emission
from the electrosphere may be of the same order as the black body radiation. 
For this reason the situation with distinguishing a bare quark star 
made of the SQM in normal (or 2SC) phase from a neutron star
using the luminosity \cite{Usov1,Usov-LC} 
may be more optimistic
than in the scenario with the tunnel $e^{+}e^{-}$ creation \cite{Usov1}.

\vspace{.1cm}
\noindent {\bf 2}.
As in \cite{Usov1,Gale,Zh1} we use for the electrosphere the model of 
a relativistic strongly degenerate electron gas in the Thomas-Fermi
approximation. Then the electron chemical potential 
(related to the electrostatic potential, $V$, as $\mu=eV$)
may be written as
\cite{Alcock1,Kettner} 
\beq
\mu(h)=\frac{\mu(0)}{(1+h/H)}\,,
\label{eq:10}
\eeq
where
$h$ is the distance from the quark surface, 
and
$H=\sqrt{{3\pi}/{2\alpha}}/\mu(0)$, 
$\alpha=e^{2}/4\pi$
(we use units $c=\hbar=k_{B}=1$).

We assume that the electrosphere is optically thin. Then the luminosity
may be expressed in terms of the energy  radiated
spontaneously per unit time and volume, $Q_{\gamma}$,
usually called the emissitivity.
In the formalism \cite{AZ} the emissitivity per unit photon energy $\omega$ 
at a given $h$  can be written as 
\beq
\frac{dQ_{\gamma}(h,\omega)}{d\omega}=
\frac{\omega(k)}{4\pi^{3}}
\frac{dk}{d\omega} 
\int \frac{d\pb}{p} n_{F}(E)[1-n_{F}(E')]\theta(p-k)
\frac{dP(\pb,x)}{dx dL}\,,
\label{eq:20}
\eeq
where 
$k$ is the photon momentum, 
$E$ and $E'$ are the electron energies 
before and after photon emission, 
$n_{F}(E)=(\exp((E-\mu)/T)+1)^{-1}$ is the local electron Fermi
distribution (we omit the argument $h$ in the functions 
on the right-hand side 
of (\ref{eq:20})),
$x=k/p$ is the photon longitudinal (along the initial electron momentum
$\pb$) fractional  
momentum. The function $dP/dx dL$ in (\ref{eq:20}) is the probability  
of the photon emission per unit $x$ and length from an electron in 
the potential generated by other electrons which includes both the 
smooth collective Coulomb field
and the usual fluctuating part.
Note that (\ref{eq:20}) assumes that 
the photon emission is a local process, {\it i.e.} the photon 
formation length $l_{f}$ is small compared to the thickness of the 
electrosphere. 

In the LCPI formalism \cite{Z1,Z_YAF98} 
the photon spectrum $dP/dx dL$
can be written as 
\beq
\frac{d P}{d
xdL}=2\mbox{Re}
\int\limits_{0}^{\infty}d
\xi
%\left.
\hat{g}(x)\left[{\cal
K}(\r',\xi|\r,0)
-{\cal
K}_{v}(\r',\xi|\r,0)
\right]
%\,\right
\Big{|}_{\r^{'}=\r=0}.
\label{eq:30}
\eeq
Here $\hat g$ is the spin vertex operator 
(it can be found in \cite{Z_YAF98}),
${\cal K}$ is the Green's function for the 
two-dimensional Hamiltonian
\beq
\hat{H}=-\frac{1}{2M(x)}
\left(\frac{\partial}{\partial \r}\right)^{2}
+v(\r)+\frac{1}{L_{0}}\,,
\label{eq:40}
\eeq
where $M(x)=px(1-x)$,  
$L_{0}=2M(x)/\epsilon^{2}$,
$\epsilon^{2}=m_{e}^{2}x^{2}+(1-x)m_{\gamma}^{2}$ 
($m_{\gamma}$ is the photon quasiparticle mass),
the form of the potential $v$ 
will be given below.
In (\ref{eq:30}),~(\ref{eq:40}) $\r$ is the coordinate transverse to the 
electron momentum $\pb$, the longitudinal (along $\pb$) coordinate $\xi$
plays the role of time.
The ${\cal K}_{v}$ in (\ref{eq:30}) is the free Green's function for $v=0$.
Note that at low density and vanishing mean field the quantity $L_{0}$ 
coincides with the real photon formation length $l_{f}$ \cite{Z1}.

The potential in the Hamiltonian (\ref{eq:40}) 
can be written as $v=v_{m}+v_{f}$. The terms  
$v_{m}$ and $v_{f}$ correspond to the mean and fluctuating
components of the vector potential of the electron gas.
Note that when $l_{f}$ is small compared to the scale of variation 
of $\mu$ (along the electron momentum)
one can neglect the $\xi$-dependence of the potential $v$
in evaluating $dP/dxdL$. 
The mean field component is purely real
$v_{m}=-x\fb\!\cdot\!\r$ with $\fb=e \partial V/\partial \r$ 
\cite{Z_YAF98,Z_synch}. 
It is related to the transverse force from the mean field.
Note that, 
similarly to the classical radiation 
\cite{LL2}, the effect of the longitudinal force 
along the electron momentum $\pb$ is suppressed by a factor 
$\sim (m_{e}/E)^{2}$, and can be safely neglected.
The term $v_{f}$ 
can be evaluated similarly to the case 
of the quark-gluon plasma discussed in \cite{AZ}.  
This part
is purely imaginary 
$v_{f}(\r)=-i P(x\r)$, where
\beq
P(\r)=e^{2}\int\limits_{-\infty}^{\infty} d\xi 
[G(\xi,0_{\perp},\xi)-G(\xi,\r,\xi)]\,,
\label{eq:50}
\eeq
$
G(x-y)= 
u_{\mu}u_{\nu} D^{\mu\nu}
$,
$
D^{\mu\nu}={\Large\langle}
A^{\mu}(x)A^{\nu}(y)
{\Large\rangle}
$
is the correlation function of the electromagnetic potential
(the mean field is assumed to be subtracted)
in the electron plasma, $u_{\mu}=(1,0,0,-1)$ is the light-cone 4-vector
(along the electron momentum).
The correlator $D^{\mu\nu}$ may be expressed in 
terms 
of the longitudinal and transverse  photon self-energies, $\Pi_{L,T}$
\cite{AMY1}.
In numerical calculations we use for the $\Pi_{L,T}$ the well known 
hard dense loop expressions \cite{HDL1,HDL2}.

Treating $v_{f}$ as a perturbation one can write
\beq
{\cal K}(\xi_2,\r_{2}|\xi_1,\r_{1})=
{\cal K}_{m}(\xi_2,\r_{2}|\xi_1,\r_{1})
-i\int d\xi d\r
{\cal K}_{m}
(\xi_2,\r_{2}|\xi,\r)
v_{f}(\r){\cal K}_{m}(\xi,\r|\xi_1,\r_{1})+\dots\,\,,
\label{eq:100}
\eeq
where ${\cal K}_{m}$ is the Green's function for $v_{f}=0$.
Then (\ref{eq:30}) can be written as
\beq
\frac{d P}{d
xdL}=\frac{d P_{m}}{d
xdL}+
\frac{d P_{f}}{d
xdL}\,.
\label{eq:110}
\eeq
Here the first term on the right-hand side comes from the
${\cal K}_{m}-{\cal K}_{v}$ in (\ref{eq:30}) after representing
${\cal K}$
in the form (\ref{eq:100}). It corresponds to the photon emission
in a smooth mean field.  The second term 
comes from the series in $v_{f}$ in (\ref{eq:100}). This term can be viewed
as the radiation rate due to electron multiple scattering in the fluctuating
field in the presence of a smooth external field.
The analytical expression for the Green's function ${\cal K}_{m}$
is known (see, for example \cite{FH}). The corresponding spectrum
is similar to the well known synchrotron spectrum, and 
can be written in terms of the Airy function
$\mbox{Ai}(z)=\frac{1}{\pi}\sqrt{\frac{z}{3}}K_{1/3}(2z^{3/2}/3)$ (here
$K_{1/3}$ is the Bessel function) 
\cite{Z_synch,BK}. In the case of interest, for a nonzero photon
quasiparticle mass it reads \cite{Z_synch}
\beq
\frac{dP_{m}}{dxdL}=
\frac{a}{\kappa}\mbox{Ai}^{'}(\kappa)+
b\int_{\kappa}^{\infty}dy\mbox{Ai}(y)\,,
\label{eq:120}
\eeq
where 
$a=-{2\epsilon^{2}g_{1}}/{M}$,
$
b=M  g_{2}-{\epsilon^{2}g_{1}}/{M}
$,
$\kappa=\epsilon^{2}/(M^{2}x^{2}\fb^{2})^{1/3}$,
$g_{1}=\alpha(1-x+x^{2}/2)/x$ and
$g_{2}=\alpha m_{e}^{2}x^{3}/2M^{2}$.
Note that the effective photon formation length for the 
mean field mechanism is given by 
$\bar{L}_{m}\sim \mbox{min}(L_{0},L_{m})$, where 
$L_{m}=(24M/x^{2}\fb^{2})^{1/3}$ \cite{Z_synch}.

Evaluation of the ${dP_{f}}/{dxdL}$ for realistic function $P(\r)$ and
nonzero mean field is a complicated computational problem.
In the present work we have performed a qualitative calculation
of this term. We evaluated ${dP_{f}}/{dxdL}$ for zero mean field
within the LCPI formalism \cite{Z1} using the method of 
\cite{Z_SLAC1,Z_SLAC2}. 
This calculations show that for zero mean field the spectrum is dominated
by the leading order term in $v_{f}$ on the right-hand side of (\ref{eq:100}),
and the effect of the higher order terms that describe the 
Landau-Pomeranchuk-Migdal (LPM) suppression is negligible
\footnote{One can show that a very strong 
LPM suppression obtained in \cite{Gale} is due to 
use of Migdal's formulas for ordinary materials 
which become inadequate for the electrosphere.}. 
The mean field should suppress the radiation rate.
Qualitatively the corresponding suppression factor can be written
as the ratio of the formation lengths with and without
the mean field, {\it i.e.} $S_{m}\approx\bar{L}_{m}/L_{0}$.
Note that due to reduction of the effective formation length 
the LPM effect  should become even smaller for
a nonzero mean field. 
As will be seen from our numerical results 
the fluctuation term in (\ref{eq:110}) is much
smaller than the mean field one. 
For this reason getting of an accurate prediction for 
${dP_{f}}/{dxdL}$ is not important in a pragmatical sense.
Note that, since the mean field mechanism dominates, the $l_{f}$
is simply given by $\bar{L}_{m}$. 

\vspace{.1cm}
\noindent {\bf 3}.
In numerical calculations we define  
the $k$-dependent
photon quasiparticle mass from the relation
$m^{2}_{\gamma}=\Pi_{T}(\sqrt{k^{2}+m^{2}_{\gamma}},k)$.
This gives
$m_{\gamma}$ rising from $m_{D}/\sqrt{3}$ at $k\ll m_{D}$ to $m_{D}/\sqrt{2}$
at $k\gg m_{D}$
with the Debye mass 
%given by
$m_{D}^{2}=\frac{4\alpha}{\pi}(\mu^{2}+\pi^{2}T^{2}/3)$.
We ignore the influence of the medium
effects on $m_{e}$ \cite{me} since the results 
are not very sensitive to the electron quasiparticle mass.

As we mentioned earlier, the collinear approximation we use   
becomes invalid
for very soft photons with $k\lsim m_{\gamma}$. 
In this region the formalisms \cite{AMY1,AZ,Z1} do not apply.
In particular, the LCPI approach \cite{Z1},
which assumes that the transverse momentum integration
comes up to infinity, should overestimate the photon spectrum at
$k\lsim m_{\gamma}$.
To take into account (at least, qualitatively)  this effect 
we multiplied $dP/dxdL$ by the kinematical suppression factor 
$S_{kin}(k)=1-\exp(-k^{2}/m_{\gamma}^{2})$. This factor
suppresses the luminosity by $\sim 10-15$\%
at $T\sim 0.1\div 0.2$ MeV and  
$\sim 1-2$\%
at $T\sim 1$ MeV. This says that the errors from
the noncollinear configurations
are small.

We evaluated the differential,
$dF/d\omega$, and the total energy flux, $F$.
In our approach 
(the approximation of optically thin 
electrosphere)
the $dF/d\omega$ reads
\beq
\frac{dF}{d\omega}=\int_{0}^{h_{max}}dh 
\frac{d Q_{\gamma}(h,\omega)}{d\omega}
\approx
\sqrt{\frac{3\pi}{2\alpha}}
\int_{\mu_{min}}^{\mu(0)}\frac{d\mu}{\mu^{2}}
\frac{dQ_{\gamma}(h(\mu),\omega)}{d\omega}
\label{eq:160}
\eeq
with $\mu_{min}=\mu(h_{max})$.
We take 
$\mu_{min}=2m_{e}$. Of course, the relativistic approximation we made 
is not good at  $\mu\sim m_{e}$. However, the 
contribution of this region is small, and the errors  should not be big.
We have performed computations for 
$\mu(0)=10$ and $\mu(0)=20$ MeV.
In Fig.~1 we plot the radiation rate $dF/d\omega$ for the mean
field and the fluctuation mechanisms for $T=0.2$ and $T=1$ MeV.
For comparison the black body
spectrum is also shown. 
For the fluctuation contribution we show the results with and without
the mean field suppression factor $S_{m}$. One can see that the 
Coulomb potential
of the electrosphere reduces the fluctuation term by a factor $\sim 3-4$.
From Fig.~1 one can see that the relative contribution
of the fluctuation mechanism is very small. Thus, in some sense
we have a situation similar to that for an atom with large $Z$.
Note that the form of the spectrum for the 
mean field mechanism is qualitatively similar to that 
for the black body radiation.

In Fig.~2 we show the total energy flux 
$
F=\int_{0}^{\infty}d\omega {dF}/{d\omega}
$
scaled to the black body limit
as a function of temperature. 
For comparison, in Fig.~2 we also plot the energy flux from
the $e^{+}e^{-}$ pair production \cite{Usov1,Usov2}. We define it 
as
\beq
F_{e^{+}e^{-}}=\int_{0}^{h_{max}}dh Q_{e^{+}e^{-}}(h)\approx
\sqrt{\frac{3\pi}{2\alpha}}\int_{\mu_{min}}^{\mu(0)}\frac{d\mu}{\mu^{2}}
Q_{e^{+}e^{-}}(h(\mu))\,.
\label{eq:170}
\eeq
Here $Q_{e^{+}e^{-}}$ is the energy flux from $e^{+}e^{-}$ pairs 
per unit time and volume. We write it in the form given in \cite{Usov2} 
$
Q_{e^{+}e^{-}}=2E_{e^{+}e^{-}}dN_{e^{+}e^{-}}/dtdV
$, where $E_{e^{+}e^{-}}\approx 2(m_{e}+T)$ is the typical energy of 
$e^{+}e^{-}$ pairs, and $dN_{e^{+}e^{-}}/dtdV$ is the rate of 
$e^{+}e^{-}$ pair production per unit time and volume defined 
by the formulas given in \cite{Usov2}.
From Fig.~2 one sees that in the region $T\sim 0.1\div 1$ MeV the mean 
field photon emission exceeds considerably both the 
fluctuation bremsstrahlung and the energy flux from $e^{+}e^{-}$ pair
production.

 As we mentioned earlier,
our assumption that the photon emission is a local process
is valid if $l_{f}\sim \bar{L}_{m}\ll L_{el}$, 
where $ L_{el}$ is the typical scale of
variation of the potential $v_{m}$ along the electron trajectory. 
For the chemical potential (\ref{eq:10})
it can evidently be defined as $L_{el}\sim H\mu(0)/\mu(h) \cos{\theta}$,
where $\theta$ is the angle between the electron momentum and the star surface
normal. Evidently the contribution of the configurations
with $\bar{L}_{m}\gsim L_{el}$ into the photon spectrum 
will be suppressed by the finite-size suppression factor $S_{fs}\sim
\mbox{min}(L_{el},\bar{L}_{m})/\bar{L}_{m}$. We have checked numerically 
that this suppression factor gives a negligible effect. This justifies
the local approximation.

Figs.~1,~2  demonstrate that the energy flux from the mean 
field photon emission
may be of the same order of magnitude as the black body radiation.
It says that the approximation of optically thin electrosphere is not
very good, and the photon absorption and stimulated emission may be important.
However, since the radiation rate we obtained does not exceed 
the black body limit,  
they  cannot modify strongly our
results \footnote{The authors of \cite{Harko2} obtained for 
$\mu(0)\sim 10-20$ MeV and $T\lsim 1$ MeV the energy flux
considerably exceeding the black body limit. They claim
that it is possible for the electrosphere. This statement is obviously 
incorrect. The violation of the black body limit in 
\cite{Harko2} is just a signal that the thin medium approximation 
becomes inadequate at high emissitivity. As far as a very large emissitivity
obtained in \cite{Harko2} is concerned, as we already mentioned, 
it may be due to incorrect description
of the Pauli-blocking and neglect of the photon mass.
}.

According to simulation of the thermal evolution of young
quark stars performed in \cite{Usov-LC} the temperature at the 
star's surface becomes $\sim 0.2$ MeV at $t\sim 1$ s. 
However, in the analysis \cite{Usov-LC} the mean field bremsstrahlung
was not taken into account. 
In the light of our results
one can expect that the cooling of the bare quark star's surface  
should go somewhat 
faster than predicted in 
\cite{Usov-LC}
\footnote{It is worth noting, however, that in the initial hot stage
the mean field bremsstrahlung will 
change only the temperature of the quark star near its surface. 
While the evolution of the star core 
temperature is driven by the neutrino emission \cite{Usov-LC}
since for an extended period of time the neutrino luminosity is much 
larger than the photon (and
$e^{+}e^{-}$) luminosity \cite{Usov-LC}.}.
Higher luminosity 
due to the mean field bremsstrahlung
increases the possibility for detecting bare quark stars.   
From the point of view of the light curves at $t\gsim 1$ s it would 
be interesting to investigate the mean field bremsstrahlung
for $T\lsim 0.1$ MeV as well. 
However, at such temperatures the photon emission from the
nonrelativistic region of the electrosphere may be important,
where our formulas become inapplicable. As far as the contribution of the
relativistic region $\mu\gg m_{e}$ is concerned. Extrapolation of the
curves shown in Fig.~2 to 
$T\lsim 0.1$ MeV allows one to expect that 
the mean field emission will dominate the energy flux at lower temperatures
as well. However, a robust conclusion on the relative contributions
of the photon emission and $e^{+}e^{-}$ pair production can only be made
after calculating the photon bremsstrahlung beyond the collinear approximation
(in the relativistic and nonrelativistic regions of the electrosphere).

It is worth noting that for $T\sim 0.1\div 1$ the form of the differential
radiated energy flux and the relative fractions of photons and $e^{\pm}$
pairs are not important from the point of view of the photon
spectrum observed at large distances from the star. One can show
that in this temperature region for the energy flux of the 
order of the black body limit the outflowing wind of photons, electrons and 
positrons is thermalized at distances much smaller than the star radius.
For the thermalized $e^{\pm}\gamma$ wind the photon distribution seen 
by a distant observer is close to the black body one, and 
the fraction of electrons and positrons is negligible \cite{Pacz}.
For this reason the specific form of the photon spectrum from 
$e^{+}e^{-}$ annihilation for the tunnel 
$e^{+}e^{-}$ creation mechanism \cite{Usov1} is not important in
the investigated temperature window. It may be important only at much 
smaller temperatures in the regime of a free streaming $e^{+}e^{-}\gamma$
wind.

The calculations of the photon emission from bare quark stars
in the color flavor locked (CFL) superconducting phase 
(when electrons are probably absent even near the star surface
\cite{CFL-el}) performed in \cite{Rapp} give the radiation
rate comparable to the black body limit.
Since we also obtain the radiation rate comparable to the black body 
radiation it may be difficult to distinguish a
bare quark star in the CFL phase from that in normal 
(or 2SC) phase.

\vspace{.1cm}
\noindent {\bf 4}.
In summary, using the LCPI reformulation \cite{AZ} of the AMY 
approach \cite{AMY1} to the photon emission from relativistic plasmas 
we have calculated 
the photon emission from
the electrosphere of a
bare quark star (in normal or 2SC phase).
Contrary to the previous qualitative studies \cite{Gale,Zh1,Harko2},
it allows, for the first time, to give a robust treatment
of the Pauli-blocking effects in the photon bremsstrahlung.
We demonstrate that for the temperatures $T\sim 0.1\div 1$ MeV
the dominating contribution to the photon
emission is due to bending of electron trajectories 
in the mean electric field of the electrosphere.
The energy flux from the mean field photon emission
is of order of the black body limit.
Our results show that the contribution 
of the bremsstrahlung due to electron-electron interaction
is negligible as compared to the mean field photon emission.

The energy flux related to the mean field bremsstrahlung
turns out to be larger than that from the tunnel $e^{+}e^{-}$ creation
\cite{Usov1,Usov2} as well.
In the light of these results the situation with distinguishing 
bare quark stars made of the SQM in normal (or 2SC) phase 
from neutron stars may be more optimistic
than in the scenario with the tunnel $e^{+}e^{-}$ creation discussed
in \cite{Usov-LC}. 

\vspace {.7 cm}
\noindent
{\large\bf Acknowledgements}

\noindent
I am grateful to T. Harko and D. Page for communication. 
This work is supported in part by the grant SS-6501.2010.2.

%\pagebreak

\newpage
%------------------------------------------------------------------
\begin{figure}[htb]%[h]
\begin{center}
\epsfig{file=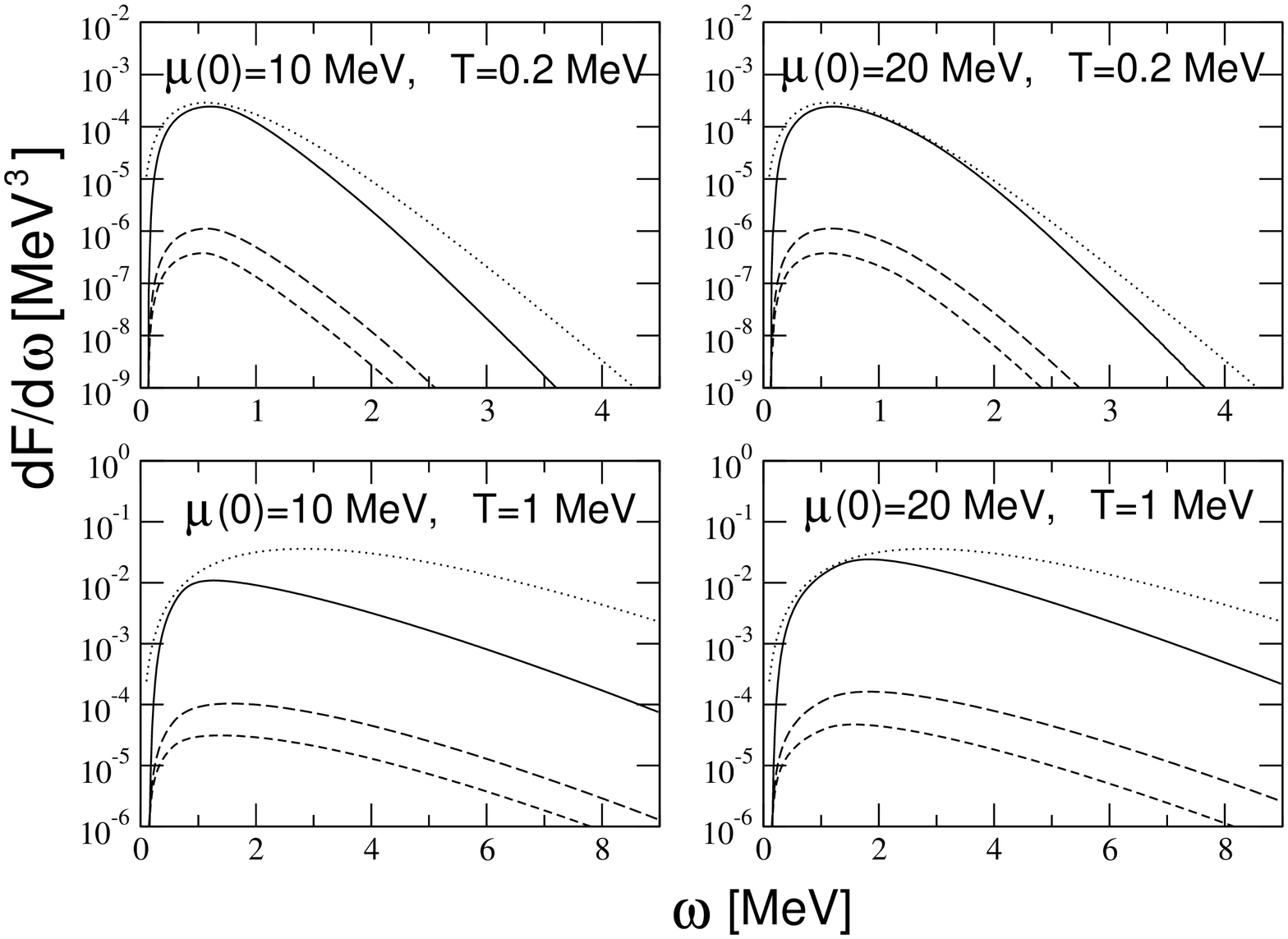,height=13.7cm}
\end{center}
\caption[.]{
The differential radiated energy fluxes from the electrosphere
for the mean field
bremsstrahlung (solid line) and for the bremsstrahlung due to 
electron-electron interaction with (short dashes) and 
without (long dashes) the mean field suppression.
%factor $S_{m}$.% described in the text. 
The dotted curves show the black body
spectrum.
}
\end{figure}
\begin{figure}[htb] %[h]
\begin{center}
\epsfig{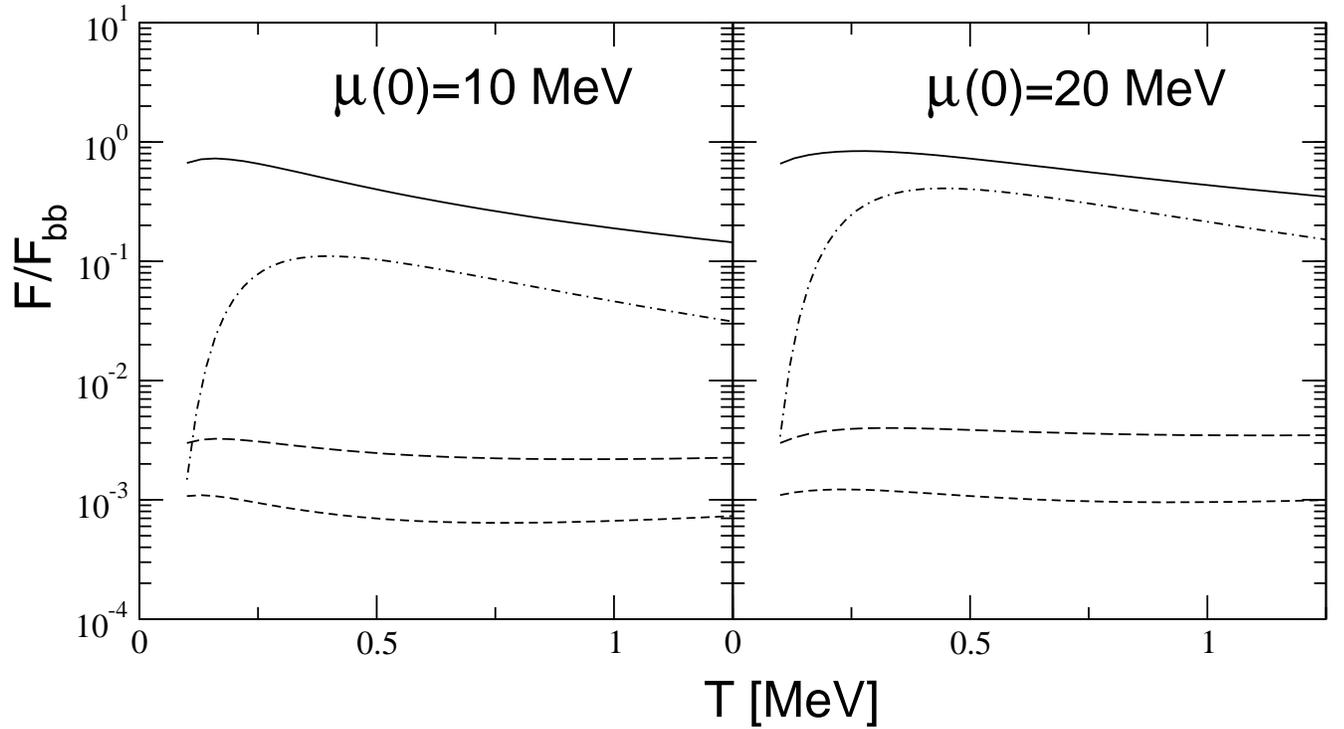}
\end{center}
\caption[.]{
The total radiated energy fluxes 
(scaled to the black body radiation)
from the electrosphere
for the mean field
bremsstrahlung (solid line) and for the bremsstrahlung due to 
electron-electron interaction with (short dashes) and 
without (long dashes) the mean field suppression.
The contribution from 
the tunnel $e^{+}e^{-}$ creation \cite{Usov1,Usov2}
evaluated using (\ref{eq:170}) is also shown (dash-dotted line).
}
\end{figure}

\end{document}